# Of the People, By the Algorithm: How AI Transforms Democratic Representation

Yuval Rymon, Tel-Aviv University[1]

yuvalrymon@mail.tau.ac.il

February 2025

## Abstract

*This review examines how AI technologies are transforming democratic representation, focusing on citizen participation and algorithmic decision-making. The analysis reveals that AI technologies are reshaping democratic processes in fundamental ways: enabling mass-scale deliberation, changing how citizens access and engage with political information, and transforming how representatives make and implement decisions. While AI offers unprecedented opportunities for enhancing democratic participation and governance efficiency, it also presents significant challenges to democratic legitimacy and accountability. Social media platforms' AI-driven algorithms currently mediate much political discourse, creating concerns about information manipulation and privacy. Large Language Models introduce both epistemic challenges and potential tools for improving democratic dialogue. The emergence of Mass Online Deliberation platforms suggests possibilities for scaling up meaningful citizen participation, while Algorithmic Decision-Making systems promise more efficient policy implementation but face limitations in handling complex political trade-offs. As these systems become prevalent, representatives may assume the role of architects of automated decision frameworks, responsible for guiding the translation of politically contested concepts into technical parameters and metrics. Advanced deliberation platforms offering real-time insights into citizen preferences will challenge traditional representative independence and discretion to interpret public will. The institutional integration of these participation mechanisms requires frameworks that balance the benefits with democratic stability through hybrid systems weighting different forms of democratic expression.*

**Key Words:** Democracy, Artificial Intelligence, Representation, Citizen Participation, Algorithmic Decision-Making

---

[1] An M.A. student in Political Science with a focus on Technology

# 1. Introduction

Artificial intelligence (AI) is fundamentally transforming democratic processes, from citizens' engagement with political information to their participation in decision-making and the very nature of governance and representation. For a long time, governments have increasingly adopted digital technologies to enhance public services and citizen participation in policy-making (Duberry, 2022, 1). However, AI represents a distinctive shift through its capacity for autonomous operation, self-improvement, and execution of complex tasks previously exclusive to humans (Filippucci et al., 2024, 23).

The scope of AI's potential impact on governance stems from its comprehensive capabilities. Leading institutions like the OECD and EU AI Act, define AI as machine-based systems capable of generating various outputs – from predictions and recommendations to content and decisions – that can influence both physical and virtual environments (Filippucci et al., 2024, 8). Following established "scaling laws" from the past decade, AI capabilities continue to expand as model complexity and training data increase, suggesting eventual dominance across many more human tasks (Korinek, 2024, 5-6).

This technological transformation intersects critically with modern liberal democracies' foundational mechanisms of representative governance. The emergence of large generative models capable of producing human-like text, images, and audio-visual content holds particular significance for democratic processes that rely primarily on linguistic exchange (Summerfield et al., 2024, 2). As AI introduces new possibilities for enhancing or potentially disrupting established democratic mechanisms, a fundamental question emerges: How will political representation in democracies evolve?

This literature review examines AI's transformation of democratic representation through two primary lenses: citizen participation and algorithmic decision-making, analyzing both immediate impacts and potential future changes to representatives' roles.

# 2. Citizen Participation

**Participation Changes Form**

Liberal democracies face declining rates of traditional citizen participation like voting and party membership, with this downward trend most pronounced among low-income and less educated populations over the past two decades (Duberry, 2022, 56-57). However, rather than indicating reduced political engagement overall, this shift reflects an evolution in how citizens participate politically.

Contemporary political engagement increasingly manifests through alternative channels, notably through social movements that can rapidly achieve national or global reach. Social media platforms have become crucial facilitators of this transformation, enabling citizen expression and movement coordination at unprecedented scales (Duberry, 2022, 58).

This evolution coincides with the rise of digital democratic tools, including online campaign platforms, hashtag activism, and civic tech initiatives. While these mechanisms enable more direct political participation, they often lack strong connections to formal decision-making institutions, and tend to foster issue-specific, short-term engagement rather than the sustained involvement characteristic of traditional party membership (Berg & Hofmann, 2021, 11-12).

**AI-driven Political Conversation, Sponsored by Social Media**

Information access and distribution are fundamental to policy-making and democratic representation, with those controlling these flows gaining significant advantages in problem identification and policy development (Duberry, 2022, 61). AI is situated at the center of this information ecosystem, fundamentally reshaping citizen-representative interactions.

As political participation has evolved, social media platforms and their AI algorithms have become crucial democratic intermediaries, functioning as gatekeepers that determine content newsworthiness, rank information, and filter access to content. In many countries, social media has become the predominant source for citizens to both consume and publish information, with machine learning algorithms significantly shaping the political debate (Duberry, 2022, 81, 86).

The AI-powered architecture of these platforms generates several concerning phenomena. First, algorithms prioritize viral, emotionally-charged content, creating feedback loops that favor sensationalism and extreme views over analytical discourse (Duberry, 2022, 79). Second, AI-driven "filter bubbles" restrict information access based on user preferences, creating echo chambers that reinforce existing beliefs and accelerate misinformation spread while undermining exposure to diverse perspectives. The prioritization of engagement metrics undermines citizens' access to the plurality of sources necessary for informed democratic participation (Duberry, 2022, 80-83).

Another significant challenge is the platforms' surveillance capabilities. Current machine learning algorithms embedded in social media platforms collect and analyze user data across devices and platforms. This enables unprecedented political monitoring of behaviors and preferences in real-time, allowing increasingly sophisticated personal political persuasion, with tailored messages, potentially engineering public opinion (Duberry, 2022, 98-99, 104, 126, 133).

The resulting dependency of political actors on tech companies for citizen engagement data creates a complex dynamic. Governments simultaneously attempt to regulate these platforms while relying on their data access and surveillance capabilities, creating a problematic symbiosis that challenges public-private boundaries and privacy rights (Duberry, 2022, 149; Berg & Hofmann, 2021, 13-15).

**Large Language Models, Larger Implications**

Large Language Models (LLMs) present both significant challenges and opportunities for political communication.

Summerfield et al. define epistemic challenges as phenomena which impact citizens' ability to make informed choices about representatives and policies. On the epistemic front, LLMs can generate misleading content, including sophisticated deepfake media, potentially eroding public trust in all content—creating a "liar's dividend" where even legitimate information becomes suspect. While current generic LLMs typically provide balanced responses, concerns exist about AI personalization reinforcing existing views and contributing to polarization. Their capacity for political microtargeting through preference inference and tailored messaging raises additional concerns about voter autonomy (Summerfield et al., 2024, 4-8).

However, LLMs also offer several democratic benefits. They can serve as defensive tools against misinformation through automated fact-checking and source verification. They can enhance political communication by helping representatives articulate ideas more effectively and enabling voters to better understand political issues, identify aligned parties, and learn about their rights. With expanded context capabilities, LLMs could process and analyze viewpoints from mass public discussions, providing governments with nuanced insights into potential reactions to political decisions (Summerfield et al., 2024, 4-8, 10-11).

Furthermore, LLMs show promise in improving democratic discourse quality by moderating heated political messages and suggesting less adversarial phrasings, with studies indicating users often accept such LLM-suggested reformulations. They have also demonstrated effectiveness in generating consensus statements that achieve broader group endorsement than human-written alternatives. However, these applications require careful oversight to ensure fair representation and avoid potential biases (Summerfield et al., 2024, 6-7).

**From Mini-Publics to Mass Online Deliberation**

Deliberative democracy emerged as an alternative to purely aggregative democratic approaches, emphasizing legitimacy through public reasoning among equal citizens rather than simple preference counting. However, this creates a fundamental tension: while democratic deliberation should include all citizens equally, quality deliberation traditionally only works in small groups ("mini-publics"), forcing delegation that compromises democratic legitimacy (Landemore, 2024, 1-4).

AI technologies now challenge this limitation through Mass Online Deliberation (MOD), enabling hundreds of thousands to participate in integrated virtual deliberative spaces with AI-augmented processes (Landemore, 2024, 12). MOD can be implemented through various approaches: ad-hoc for specific questions, regular interval, continuous, or interconnected processes that feed into each other (Ovadia, 2023, 3). Traditionally, Jürgen Habermas proposed a practical "two-track" deliberative model: a formal track of structured political institutions for decision-making, and an informal track of civil society discourse for agenda-setting, connected by media and political parties. MOD potentially creates a "third track" where algorithms facilitate mass deliberation with possible institutional integration (Landemore, 2024, 3-7, 12, 17-20, 28-32).

AI facilitates MOD through several key functions: real-time translation with cultural context awareness (Ovadia, 2023, 3); consensus identification through collective-response systems like Polis, successfully deployed in Taiwan, Singapore, and Finland (Small et al., 2023, 3); generation of potential consensus points, with one research showing reduced divisiveness in 65% of cases and unanimous support in 40% of previously divisive situations (Bakker et al., 2024, 1-2); and deliberation quality enhancement through exchange monitoring, comment routing, and opinion summarization (Landemore, 2024, 30-31; Summerfield et al., 2024, 6-7).

Although MOD may not achieve fully inclusive deliberation of all with all, it offers a practical approximation that maintains deliberative democracy's essential elements while potentially attracting broader participation through its facilitation that reduces cognitive burdens. Crucially, this model goes beyond simple preference aggregation by requiring genuine exchange of reasons and arguments (Landemore, 2024, 4-5, 19, 33).

Significant questions remain about MOD's integration with existing representative democracy. While it is presented as an "augmentation" to current systems, unclear issues include representatives' roles in implementing deliberative decisions, conflict resolution between deliberative and representative bodies, and necessary constitutional evolution (Landemore, 2024, 3, 32-33).

**Digital Platforms as Public Infrastructure**

Despite this optimistic vision, access barriers persist in digital democratic participation, from basic internet connectivity to digital literacy. Even connected citizens often lack understanding of how platforms use their data, while AI-powered systems suffer from transparency issues and security vulnerabilities. Hidden power structures can design these processes to serve their interests (Duberry, 2022, 201-202).

The 2019 French Yellow Vests crisis illustrates how platform design shapes democratic outcomes, as both the government and the protestors used the same AI-powered technology (Cap Collectif). The government's platform ("Grand Débat") restricted discussion to four pre-selected themes with a restrictive questionnaire format, offered no debate functionality, and avoided showing proposal popularity. The protestors' platform ("Vrai Débat") enabled broader engagement across eight themes, required structured proposals with explanations and sources, enabled voting and argumentation, and displayed popularity rankings. These contrasting approaches demonstrated how procedural rules can significantly influence deliberative outcomes (Duberry, 2022, 203, 206-208).

The current dominance of profit-driven social media platforms in digital communications (Risse, 2023, 63-65) suggests the need for what Zuckerman terms "Digital Public

Infrastructure" (DPI) – civic-oriented digital spaces prioritizing informed dialogue over engagement metrics. Just as governments develop physical infrastructure, they might create public digital communications infrastructure rather than relying on commercial entities to host civic discourse (Zuckerman, 2020).

Such DPI could enable continuous citizen participation throughout the policy cycle, from problem identification to evaluation (Duberry, 2022, 198). Administered by politically independent agencies, it could address current limitations in access, transparency, security, and design fairness. It could also allow to move beyond the current temporary and issue-specific citizen engagement that government agencies offer, toward comprehensive democratic participation.

## 3. Algorithmic Decision-Making

AI is transforming political decision-making in two distinct ways: enhancing representatives' decision-making capabilities and enabling automated decision processes.

Large Language Models (LLMs) augment representatives' capabilities by processing and summarizing vast data sets, supporting policy development, and assisting in legislative drafting. This enables more effective bill writing and debate, better detection of legal loopholes, and creation of precise "micro legislation" – targeted text changes that affect law implementation (Summerfield et al., 2024, 10-11).

Meanwhile, Algorithmic Decision Making (ADM) systems offer automation potential for certain political and administrative decisions by processing data, identifying patterns, and making recommendations based on predefined criteria. ADM systems require two essential conditions to operate effectively: unambiguous, stable objectives with clear success metrics, and regular decision scenarios that provide consistent patterns to learn from (König & Wenzelburger, 2022, 135, 139-141).

**Politics Doesn't Reduce to KPIs**

While ADM systems promise enhanced efficiency and consistency in decision-making, their application to political contexts faces limitations. Unlike business decisions with quantifiable outcomes and clear historical patterns, political decisions often involve novel situations or familiar issues under significantly different circumstances. Where business can reduce trade-offs to monetary terms, political decisions require balancing complex, hard-to-quantify impacts across multiple domains – as seen for example in weighing reduced crime rates through predictive policing against restrictions on civil liberties (König & Wenzelburger, 2022, 137).

This challenge becomes particularly acute with moral policy issues, which involve fundamentally incompatible values that resist reduction to optimization problems. Decisions about same-sex marriage or abortion aren't computational problems to be solved, but expressions of competing beliefs and values. Even economic policy, despite its measurable outcomes and broader consensus on goals like reducing unemployment, involves complex societal trade-offs that cannot be translated into simple algorithmic objectives (König & Wenzelburger, 2022, 136-137).

Consequently, ADM systems find their most effective application in "low politics" and administrative decisions, particularly in implementing policies after high-level goals have been democratically established. Once broad policy objectives are set through democratic processes, ADM can optimize their implementation through data analysis and pattern recognition (König & Wenzelburger, 2022, 139-141).

**Who Shapes the Algorithm Holds the Power**

Even in seemingly suitable administrative applications, ADM systems face fundamental challenges that transform them from neutral tools into contested political instruments. Their effectiveness depends critically on data quality and the translation of policy objectives into technical parameters—a process vulnerable to manipulation as political actors may disregard or distort evidence that conflicts with their goals.

This technical ambiguity in defining optimization metrics makes ADM systems another form of expert knowledge that can be leveraged to support particular political agendas. The power to design and implement these systems becomes a crucial point of political contestation, raising critical questions about control over these increasingly important governance tools (König & Wenzelburger, 2022, 143-145). Moreover, the perceived legitimacy of ADM can enhance expert influence in policy-making, potentially narrowing the scope of democratic decision-making by questioning not whether citizens are capable of self-rule, but whether they should be making decisions at all in many domains (Jungherr, 2023, 6).

**Algorithms Must Answer to the People**

ADM in policy-making faces three key legitimacy challenges: the input dimension, where citizens lack transparency and control over data and criteria used; the throughput dimension, where complex and opaque algorithmic logic (particularly in deep learning models comprised of neural networks) obscures decision-making processes, limiting their ability to challenge decisions; and the output dimension, where questions persist about ADM's ability to improve policy outcomes while maintaining democratic principles like non-discrimination (Duberry, 2022, 203, 206-208).

The "human-in-the-loop" (HITL) approach emerges as a potential solution, with Crootof, Kaminski, & Price (2023, 474-484) defining various roles humans can play in ADM. Two roles prove particularly crucial for democratic representation:

First, the accountability role establishes clear responsibility for decisions—a fundamental democratic requirement. As power flows from the people, they must be able to hold decision-makers responsible. Human accountability prevents the diffusion of responsibility into inscrutable technical systems, maintaining citizens' ability to evaluate outcomes and respond through electoral processes (Crootof, Kaminski, & Price, 2023, 482-484).

Second, the justificatory role enables explanation of decisions in comprehensible terms, crucial for legitimacy. A core aspect of legal systems' legitimacy is allowing people to understand and potentially contest decisions that affect them. In high-stakes contexts like criminal justice, the ability to understand why a decision was made can be crucial for both accepting the decision and deciding whether and how to contest it. While algorithmic systems often struggle to explain their decisions, humans can provide justifications and make outcomes understandable to affected parties. However, this role risks becoming merely "deceptively justificatory" if humans offer post-hoc rationalizations without genuine understanding of the algorithmic decision-making process (Crootof, Kaminski, & Price, 2023, 478-480).

**Adding Humans is not Enough**

The "MABA-MABA trap" (Men Are Better At - Machines Are Better At) reveals a common policy misconception: that adding humans to machine systems will automatically combine the strengths of both. However, as Crootof, Kaminski, & Price (2023, 470-472) demonstrate, hybrid systems often create new types of systemic failures, potentially performing worse than either component alone.

These hybrid failures manifest in multiple ways: human deliberation can bottleneck algorithmic efficiency, human bias can compromise algorithmic consistency, and algorithmic rigidity can impair human contextual judgment. Moreover, humans frequently "rubber-stamp" algorithmic decisions due to "automation bias," leading to excessive deference to machines and eventual "skill fade" – the loss of ability to perform or oversee automated tasks. This dynamic becomes particularly problematic since humans are asked to monitor systems that outperform them by-design (Crootof, Kaminski, & Price, 2023, 468-469). Poor interface design often exacerbates these human-machine interactions, creating unique failures that wouldn't occur with either component alone. This was tragically demonstrated in the Boeing 737 Max crashes where information transfer between system and pilots broke down (Crootof, Kaminski, & Price, 2023, 470-472).

## 4. The Future of Representation

The integration of AI into governance can potentially transform how representatives function.

In the coming years, as ADM systems with human oversight become more prevalent, representatives may assume the role of being architects of automated decision frameworks, responsible for guiding the translation of politically contested concepts like fairness and discrimination into technical-operational parameters. This role demands careful balance between algorithmic efficiency and democratic values, particularly when core principles like equality, transparency, and individual autonomy conflict with optimization potential. Additionally, representatives will need to mediate between citizen preferences and the concrete systems' parameters and oversight mechanisms. Thus, they are likely to embody the justificatory and accountability roles of HITL, as their policy will be built into the design of ADM systems.

Representatives will also need to navigate new forms of public input. Advanced deliberation platforms offering real-time, granular insights into citizen preferences may diminish traditional representative discretion, challenging their traditional role as interpreters of the public will. Representatives might need to explicitly justify deviations from clearly expressed public preferences, potentially becoming facilitators reconciling public demands with practical constraints.

The institutional integration of digital public participation presents significant opportunities and challenges for democratic legitimacy. While clear channels for public input and transparent influence procedures could enhance legitimacy, risks include participation inequality through digital divides, deliberation platform manipulation and security, and tensions between rapid feedback and thoughtful policymaking. States will need to develop frameworks balancing digital participation's benefits with democratic stability, potentially through hybrid systems that appropriately weight different forms of democratic expression across various decision contexts. The question of integration is not binary but rather about finding the right balance and the appropriate mechanisms for different contexts and types of decisions.

These changes in representative democracy will likely manifest unevenly across political systems and jurisdictions. While some regions may quickly adopt integrated digital participation and ADM systems, others will maintain traditional representative forms longer. This period of institutional experimentation could yield valuable insights for modernizing representative democracy while preserving its essential functions.

## 5. Conclusion

This literature reveals a profound potential for artificial intelligence to transform democratic representation. The research presented suggests that AI offers unprecedented tools for enhancing democratic participation, improving information flows, and creating more responsive governance. Yet, it also highlights the risks of fragmenting public discourse, concentrating power, and undermining democratic principles.

Through processes like Mass Online Deliberation (MOD), AI can enable unprecedented scales of civic engagement, allowing hundreds of thousands of citizens to participate meaningfully in political deliberation. This capability could reinvigorate democratic processes currently struggling with declining civic participation.

AI's decision-making capabilities offer another avenue for democratic transformation. Representatives could leverage AI to process vast information, draft nuanced legislation, detect policy loopholes, and conduct sophisticated impact assessments. Algorithmic Decision Making (ADM) systems show particular promise in administrative decisions with clear objectives and consistent patterns, not replacing human judgment but augmenting it.

However, significant challenges emerge. AI integration into democratic processes raises critical questions about legitimacy, accountability, and human agency. Concerns include opaque algorithmic decision-making, technological bias, and maintaining meaningful oversight with a successful interface between humans and machines. The technology's limitations in handling complex moral decisions, along with risks to privacy and potential manipulation, demand careful consideration.

Research suggests that AI will not replace representatives but reshape their role in hybrid governance models combining human judgment with technological capabilities. Importantly, AI's impact depends on intentional design, robust governance frameworks, and commitment to democratic values of deliberation, moral judgment, and collective self-determination.

Three key areas need further study. First, mechanisms for resolving conflicts between AI-facilitated and traditional deliberative bodies, and the protocols for integrating AI-generated insights into policy-making processes. Taiwan's experience with Polis can provide meaningful lessons. Second, longitudinal studies tracking how AI technologies affect citizens' political engagement, information consumption, and decision-making processes. Third, innovative approaches to maintain accountability, rights protection, and democratic representation principles in the governance of ADM in democratic systems. As AI's capabilities evolve, their impact on democratic representation remains a critical question for society.